\documentstyle[aps,preprint]{revtex}
\def\prl#1#2{{\it Phys. Rev. Lett.} {\bf #1}, #2}
\def\pla#1#2{{\it Phys. Lett. A} {\bf #1}, #2}
\def\pre#1#2{{\it Phys. Rev. E} {\bf #1}, #2}
\def\jsp#1#2{{\it J. Stat. Phys.}  {\bf #1}, #2}
\def\pra#1#2{{\it Phys. Rev. A} {\bf #1}, #2}
\def\physr#1#2{{\it Phys. Rep.} {\bf #1}, #2}
\def\physrmp#1#2{{\it Rev. Mod. Phys.} {\bf #1}, #2}

\def\europ#1#2{{\it Europhys. Lett.} {\bf #1}, #2}

\def\physd#1#2{{\it Physica D} {\bf #1}, #2}

\def\jnsc#1#2{{\it J. Nonlinear Sci.} {\bf #1} #2}
\def\anlis#1#2{{\it Ann. Isr. Phys. Soc.} {\bf #1}, #2}
\def\funct#1#2{{\it Funct. Anal. Appl.} {\bf #1}, #2}
\def\fundma#1#2{{\it Fund. Math.} {\bf #1}, #2}
\def\progth#1#2{{\it Prog. Theor. Phys.} {\bf #1}, #2}
\def\commth#1#2{{\it Comm. Math. Phys.} {\bf #1}, #2}
\def\snsa#1#2{{\it Proc. Ind. Natl. Sci. Acad.} {\bf #1} #2}
\def\iter#1#2{{\it Int. J. Bifurcation and Chaos} {\bf #1}, #2}
\def\cha#1#2{{\it Chaos } {\bf #1}, #2}
\def\advph#1#2{{\it Adv. Physics } {\bf #1}, #2}
\def\jatms#1#2{{\it  J. Atmos. Sci. } {\bf #1}, #2}
\def\F{\bf F}
\def\JF{\bf JF}
\def\X{\bf X}

\def\e{\bf e}
\def\a{\alpha}

\def\and{\& }

\def\eg{~e. g.}
\def\etl{$et ~al.$}
\def\MLE{$\Lambda$~}

\def\sna{~SNA}

\def\beq{\begin{equation}}
\def\bc{\begin{center}}
\def\ec{\end{center}}
\def\lye{Lyapunov exponent}
\def\eqn{\end{equation}\noindent}
\begin{document}
\title{Strange Nonchaotic Attractors}
\author{Awadhesh Prasad\thanks{Present address: Department
of Mathematics, Arizona State University, Tempe AZ 85287, USA},
Surendra Singh Negi, and Ramakrishna Ramaswamy}
\address{School of Physical Sciences\\ Jawaharlal Nehru University, New
Delhi 110 067, INDIA}
%\date{\today}
\maketitle
\begin{abstract}
Aperiodic dynamics which is nonchaotic is realized on Strange
Nonchaotic attractors (SNAs). Such attractors are generic in
quasiperiodically driven nonlinear systems, and like strange
attractors, are geometrically fractal. The largest Lyapunov exponent is
zero or negative: trajectories do not show exponential
sensitivity to initial conditions. In recent
years, SNAs have been seen in a number of diverse experimental
situations ranging from quasiperiodically driven mechanical or
electronic systems to plasma discharges. An important connection
is the equivalence between a quasiperiodically driven system and the
Schr\"odinger equation for a particle in a related quasiperiodic
potential, giving a correspondence between the localized states of the
quantum problem with SNAs in the related dynamical system.
In this review we  discuss the main conceptual issues in the
study of SNAs, including the different bifurcations or routes for the
creation of such attractors, the methods of characterization, and the
nature of dynamical transitions in quasiperiodically forced systems.
The variation of the Lyapunov exponent, and the qualitative and
quantitative aspects of its local fluctuation properties, has emerged as
an important means of studying fractal attractors, and this analysis
finds useful application here.  The ubiquity of such attractors, in
conjunction with their several unusual properties, suggest novel
applications.
\end{abstract}

\newpage
\tableofcontents
\newpage
\section{INTRODUCTION}

One of the most enduring paradigms in the study of dissipative
nonlinear dynamical systems has been the concept of a strange
attractor. The term `strange', introduced  by Ruelle and Takens
[1971], is used to describe a class of attractors on which the motion is
chaotic, {\it i.e.}, showing exponential sensitivity to initial
conditions [Eckmann and Ruelle, 1985].  Most known examples of strange
attractors---the Lorenz attractor [Lorenz, 1963], for
instance---also have fractal geometry, namely they are self--similar
on different spatial scales,  and further, are properly described by a
spectrum of singular measures.

Grebogi, Ott, Pelikan, and Yorke [1984] constructed dynamical systems
with attractors that were manifestly fractal, but on which the dynamics
is {\it not} chaotic. The largest Lyapunov exponent, which is a measure
of the rate of separation of trajectories with nearby initial
conditions, is either zero or negative.  At the same time, owing to the
underlying fractal structure of the attractor, the dynamics is
intrinsically {\it aperiodic}. These {\it Strange} (namely fractal)
{\it Nonchaotic Attractors} (SNAs), which have been the focus of
considerable interest from both theoretical and experimental points of
view in the past few years, form the subject of this review.

Strange\footnote{It should be pointed out that the originally intended
meaning of the word strange in the definition of strange attractors was
not restricted to the fractal geometric aspects alone; the strangeness
was used to imply chaotic dynamics as well. The usage of the term in
the context of SNAs refers to both the spatial fractal geometry and the
temporal aperiodicity.} nonchaotic attractors, although somewhat
exotic, are not all that rare. They are {\it generic} in systems where
there is quasiperiodic forcing, and are typically found in the
neighborhood of related strange chaotic attractors in parameter space,
as well as in the neighborhood of related periodic or quasiperiodic
attractors.  In a sense they represent dynamics which is intermediate
between quasiperiodic and chaotic: there is no sensitive dependence on
initial conditions, similar to motion on regular (periodic or
quasiperiodic) attractors, but the motion is aperiodic, similar to
dynamics on chaotic attractors.

It is probably simplest to describe SNAs through an example. Consider
the  modified driven pendulum equation
\beq
\ddot x + \gamma \dot x = q_1 \sin\omega_1 t+q_2\sin\omega_2 t
-(x+\beta \sin2\pi x)\label{zmb}
\eqn
which has been used to model a driven SQUID with inertia and damping
[Zhou, Moss, and Bulsara, 1992]. For $\omega_1/\omega_2$
chosen to be an irrational ratio, the driving is quasiperiodic.
Depending on the parameters $\beta, \gamma, q_1$ and $q_2$ , the
attractors of this system, shown in Fig.~1 can be either strange and
nonchaotic (Fig.~1(a)) or strange and chaotic (Fig.~1(b)).

Visually there is little to distinguish a SNA from a strange chaotic
attractor since they superficially look very similar. Dynamically,
though, there are important distinctions.  This is most clearly evident
in the behaviour of trajectories with nearby initial conditions, shown
respectively in Figs.~1(c) and 1(d). Orbits converge and eventually
coincide on the SNA, but on the chaotic attractor, they remain distinct
and separate. Note that orbits on a chaotic attractor and on a SNA are
both aperiodic. However, because the Lyapunov exponents are zero or
negative on a SNA, trajectories do not separate from each other: thus
SNA dynamics is in a sense predictable even though it is aperiodic. The
property of robust synchronization is very characteristic of SNAs and
can be utilized in a variety of applications which require aperiodicity
[Ramaswamy, 1997].

Although the initial examples were constructed explicitly and could be
considered as being somewhat artificial, SNAs are now known to be
pertinent in a variety of physically relevant situations. The first
experimental observation of a \sna~ was in a magnetoelastic ribbon
[Ditto \etl, 1990]. This versatile mechanical system
consists of a ribbon made of amorphous magnetostrictive material,
which is clamped at the base and driven by an oscillating magnetic
field. It has been extensively used to demonstrate and to
implement a number of different ideas in the study of nonlinear
dynamical systems. Quasiperiodic driving is achieved by using
two oscillating magnetic fields with irrationally related frequencies,
and the system is then modeled by a forced Duffing--like
oscillator [Heagy and Ditto, 1991] with the equation of motion
\beq
\ddot x + \gamma \dot x = x\{1 + A[R\cos(t+\phi_0) + \cos \Omega t]\}
-x^3.
\label{heagyditto}
\eqn
where $\gamma, A$ and $R$ are parameters. If the frequency $\Omega$ is
an irrational number, the system is quasiperiodically driven; in many
cases, this number is chosen to be the golden mean ratio, $(\sqrt 5
-1)/2$.  The sensitivity of this particular experiment was sufficient
to verify the existence of SNAs from the data by estimating the fractal
dimension of the underlying attractor and by observing power--law
behaviour in the scaling of the spectral distribution (see
Eq.(\ref{scaling}) below).

SNAs can also be realized in electronic circuits, close
to the transition to chaos [Yang and Bilimgut, 1997;
Kapitaniak and Chua, 1997; Murali, Venkatesan and Lakshmanan, 1999].
In the typical case, a circuit is driven by two sinusoidal
voltage sources with the frequencies irrationally related; the
appropriate equations for the experiment of Yang and Bilimgut are
(in standard notation)
\begin{eqnarray}
\frac{dv_{c}}{dt} &=&{1\over C}[ i_{L}-\bar{g}(v_{c})] \\
\frac{di_{L}}{dt} &=&{1\over L} [R i_{L} - v_{c} +{\it f_1} sin(\omega_{1}t)+f_2 sin(\omega_{2}t),
\end{eqnarray}
Here $f_1$ and $f_2$ are the amplitudes and $\omega_{1}$ and $\omega_{2}$ are
the frequencies of the two sinusoidal voltages. $\bar{g}(v_{c})$, the
nonlinear characteristic of the linear negative resistor, is given by
$\bar{g}(v_{c}) =G_{b} v_{1}+{1\over 2}(G_{a}-G_{b})(\vert v_{1}+E\vert -
\vert v_{1}-E \vert )$ and $E$ is the breakpoint voltage.

An important area where the study of SNAs finds conceptual application
is the case of quantum mechanical systems with quasiperiodic
potentials. There is an unexpected link between wave--function
localization phenomena and the related strange nonchaotic dynamics of
an auxiliary variable, which was first pointed out by
Bondeson \etl~[1985] who considered the Schr\"odinger equation
\beq
\label{schro}
-{{d^2 \psi(x)} \over{d x^2}} + \alpha V(x) \psi(x) = E \psi(x).
\eqn
Defining $\phi(x)$ via the Pr\"ufer transformation, namely
$\exp i \phi(x) =
(\psi^{\prime} + i g \psi)/(\psi^{\prime} - i g \psi)$, $g$ being a
constant, Eq.~(\ref{schro}) can be brought into the form
\beq
\label{forced}
{d\phi \over dx} = {1\over g}\{g^2 - [E-\alpha V(x)] \cos \phi + g^2 +
[E- \alpha V(x)]\}\eqn
which is similar, for appropriate choice of potential $V(x)$ and suitable
redefinition of the independent variable, to the forced
pendulum system,
\beq
\label{pend}
\dot \phi = \cos \phi + K +V_0(\cos \omega_1 t + \cos \omega_2 t).
\eqn
It is known that SNAs exist in the above system if the frequencies
$\omega_1$ and $\omega_2$ have an irrational ratio; upon varying $K$
and $V_0$, SNAs can be observed in a finite interval in parameter
space. The transformation connecting the two
systems implies that quasiperiodic driving in the classical
forced pendulum problem, Eq.~(\ref{pend}) corresponds to a
quasiperiodic potential in the
isomorphic Schr\"odinger equation, Eq.~(\ref{schro}).

In the quantum problem, which has been extensively studied,
states can be localized or extended depending on whether $\alpha$
is relatively large or small. For the localized states, it happens that
the inverse localization length is exactly the negative of the Lyapunov
exponent of orbits in the corresponding classical problem [Bondeson
\etl, 1985]: in this case, the localization length is positive, and
therefore localized states of the quantum problem correspond to {\it
attractors} in the dual (classical) dynamical system. Further
analysis shows that these attractors have fractal geometry, thus
giving the correspondence between SNAs and localized states.

Discretization of Eq.~(\ref{schro}) permits further analysis. Ketoja
and Satija [1997] have studied the Harper equation,
\beq
\label{harper}
\psi_{k+1} - \psi_{k-1} + 2 \alpha \cos 2\pi(k\omega + \theta_0) \psi_k = E
\psi_k,
\eqn
where $k$ labels the sites of a 1--dimensional lattice, $\psi_k$ is the
wave function at the site and $\omega, \theta_0$ and $\alpha$ are
parameters, while $E$ is the energy eigenvalue.  Upon transforming to
the new variable $x_k = \psi_{k-1}/\psi_k$ (this is essentially
the discrete version of the Pr\"ufer transformation, cf.
Eqs.~(\ref{schro}-\ref{forced}) above), one obtains
\begin{eqnarray}
\label{ks}x_{k+1} &=& {-1 \over{x_k - E + 2\alpha \cos 2 \pi \theta_k}}\\
\theta_{k+1} &=& \theta_k + \omega \mbox{~~~~~~~mod~} 1, \label{ks2}
\end{eqnarray}
which, for irrational $\omega$, is a quasiperiodically driven mapping
that supports quasiperiodic, chaotic and strange nonchaotic attractors
for different values of $E$ and $\alpha$. If $\alpha > 1$, eigenstates
are exponentially localized, and correspond, as for the continuous
system, to SNAs in Eq.~(\ref{ks}-\ref{ks2}) [Ketoja and Satija, 1997].
Furthermore, the absolute value of the Lyapunov exponent of the map in
Eq.~(\ref{ks}) is also the same as the inverse localization length in
the Schr\"odinger problem if $E$ happens to coincide with an
eigenvalue. For $\alpha = 1$, states are still localized, but with
power--law [Andr\'e and Aubry, 1980] rather than exponential
localization. Such states are termed {\it critical}, and the
corresponding attractors in the classical map have all
Lyapunov exponents equal to zero [Prasad \etl, 1999].

The study of SNAs thus clearly has relevance not only to dynamically
important and unusual behaviour, but also to fundamental problems in
other areas in physics.

In this review, we focus on a number of issues pertinent to the
study of SNAs. Most known examples of systems with SNAs appear to have
either quasiperiodic parametric modulation or  quasiperiodic
forcing, in the absence of which the systems support periodic or
chaotic attractors. In Sec. II we discuss the general setting within
which strange nonchaotic dynamics may be expected to occur, and
the different techniques that have been used to characterize them.
Some of the  current interest in SNAs focuses on the question of
``routes'' or ``scenarios'' for their formation. This is both
of theoretical interest, as a counterpoint to
analogous routes or scenarios for the creation of chaotic attractors,
as well as a practical question since experimental detection of SNAs
can be facilitated if the mechanisms of creation of such behaviour
are more clearly understood. Sec. III and IV are devoted to these
aspects. In Sec.~V, we discuss experimental observation of SNAs in
diverse physical systems, and their possible applications.
The review concludes with a summary in Sec.~VI.

\section{STRANGE NONCHAOTIC DYNAMICS: Occurrence and Characterization}

There are two issues that need to be satisfactorily resolved for the
study of SNAs in a dynamical system. The first is to establish the
strangeness of the attractor and its nonchaotic nature without recourse
to numerical estimates of either the fractal dimension or the Lyapunov
exponents. The second is to establish that SNAs occur in a finite
interval in parameter space, and not just at isolated points, in which
case they would merely be mathematical curiosities which would be
difficult to observe in practice. In general these issues are not
easily settled, but for particular systems some results have been
obtained [Brindley and Kapitaniak, 1991; Keller, 1996; Bezhaeva and
Oseledets, 1996].

A driven dynamical system is usually described through equations of
motion such as Eqs.~(\ref{zmb}) and (\ref{heagyditto}) above, or as a
set of coupled ordinary differential equations,
\beq
{\bf{\dot X}} = \F(\X, t).
\eqn
Through standard procedures, say by projecting the motion onto a lower
dimensional subspace, or by sampling the variables stroboscopically,
the above continuous time dynamical system can be reduced to a discrete
mapping (see for example, Ott [1994]). It is sometimes preferable to
study discrete mappings since these capture the essential features of
the dynamics and can be mathematically easier to analyse.

Most of the quasiperiodically driven systems where SNAs have
been studied are skew--product dynamical mappings of the general form
\beq
\label{function}
\X_{n+1}  =  {\bf F}(\X_n,\theta_n)
\eqn
where $\X \in \hbox{R}^k$ is a $k$-dimensional vector and $\theta \in
\hbox{S}^1$ is a scalar which varies quasiperiodically. The simplest
(but not the only) way in which this is accomplished is via the rotation,
\beq
\theta_{n+1}  =  \theta_n + \omega \mbox{~mod~} 1,
\label{irra}
\eqn
since if $\omega$ is an irrational number, then successive iterates
of $\theta$ will densely and uniformly cover the unit interval in a
quasiperiodic manner. The parameters of the mapping ${\bf F}$ are
modulated via $\theta$ (see the examples below), resulting in a
quasiperiodically
driven dynamical system. A particular case that has been extensively
studied is the driven 1--dimensional logistic map [Heagy and Hammel,
1994], namely
\beq
x_{n+1} \equiv F_{\alpha,\epsilon}(x_n,\theta_n) = \alpha [1 + \epsilon
\cos(2\pi \theta_n)] x_n (1 - x_n),
\label{lo}
\eqn
with $\omega$ usually taken to be the golden mean ratio.
The phase--diagram for this system which details the
different dynamical states that are obtained as a function of the
parameters $\alpha$ and $\epsilon$ has been extensively studied [Prasad
\etl, 1997, 1998; Witt \etl, 1997] and is shown in Fig.~2. SNAs occur
in several different parameter ranges, between regions of periodic or
torus attractors and regions of chaotic attractors.  Indeed, the
dynamics in the transition region between regular and chaotic motion is
quite complicated, and the boundaries separating different types of
attractors are very convoluted.

The system first considered by Grebogi, Ott, Pelikan, and Yorke [1984]
is a similar mapping with (cf. Eq.~(\ref{function}))
\beq
F_{\alpha}(X_n,\theta_n) \equiv 2 \alpha \cos 2 \pi \theta_n \tanh x_n.
\label{gopy} \eqn
The following arguments [Grebogi \etl, 1984] establish the existence
of SNAs for appropriate values of $\alpha$.
\begin{itemize}
\item
In the absence of the quasiperiodic driving, the mapping $
x_{n+1} = 2 \alpha \tanh x_n$ is 1--1 and contracting, mapping
the real line into the interval $[-2\alpha, 2\alpha]$.
Because $\omega$ is an irrational number, the dynamics in $\theta$ [cf.
Eq.~(\ref{irra})] is ergodic in the unit interval.  The attractor
of the dynamical system Eq.~(\ref{gopy}) therefore must be contained in the
strip  $[-2\alpha, 2\alpha] \otimes [0,1]$.

\item
A point $x_n$ with the corresponding $\theta_n = 1/4$ will map to
($x_{n+1}=0, \theta_{n+1}= \omega + 1/4$), after which subsequent
iterates will all remain on the line ($x=0, \theta$). The same holds
for points with $\theta_n = 3/4$.  This  line
therefore forms an invariant subspace: an orbit starting on
($x=0, \theta$) will stay on it.

\item
However, as can be seen by
taking the derivative of $F$, the dynamics on this line is {\it
unstable} if $\vert \alpha \vert > 1$ (the Lyapunov exponent, defined
below, can be computed exactly within the invariant
subspace, and happens to be $\log
\vert \alpha \vert$). Thus, it follows that the attractor has a dense
set of points on the line $x=0, \theta \in [0,1]$ (since $\omega$ is
irrational), but the entire line itself cannot be the attractor
for $\alpha >
1$, since the dynamics is unstable on that line.

\item
The attractor therefore must have an infinite set of discontinuities
and therefore, a fractal structure. An explicit (numerical) computation
for particular values of $\vert \alpha \vert > 1$ yields a negative
value for the nonzero Lyapunov exponent, confirming that the attractor
is both strange and nonchaotic.

\end{itemize}

Bezhaeva and Oseledets [1996] have rigorously shown for this system
that for any irrational $\omega$ and for $\alpha$ sufficiently large, a
SNA with a singular--continuous spectrum exists.  Stronger results
which can be obtained [Keller, 1996] confirm that the attractor is a
fractal and is the topological support of an ergodic SRB
(Sinai--Bowen--Ruelle) measure.  Eckmann and Ruelle [1985] discuss the
utility of such measures for fractal attractors on which the invariant
measure is spatially nonuniform.

Another class of systems where some analytical results are possible are
the strange attractors of the Harper equation. Given the correspondence
between the dynamical system and the quantum problem, it is possible to
establish the existence of attractors with a negative Lyapunov exponent
and a fractal structure [Ketoja and Satija, 1997; Prasad \etl, 1999].
Mathematically rigorous results are more difficult in this case, but
there is some indication that it will be possible to have fractal
nonchaotic attractors in a large class of systems, even in the
absence of explicit quasiperiodic driving [Negi and Ramaswamy, 2000a].

Most studies of SNAs to date have made
recourse to heuristic arguments and ``experimental'' verifications,
through explicit computation of the Lyapunov exponents and fractal
dimensions.
A number of different quantitative methods have been introduced over
the past several years that have proven useful in verifying the
strangeness of these nonchaotic attractors.

\subsection{Lyapunov exponents and
Fractal Dimensions:} For a dynamical system
of the form of Eq.(\ref{function}), there are $k$ nontrivial Lyapunov
exponents $\Lambda^m, m= 1,2,\ldots,k$. These characterize the manner
in which a $k$--dimensional parallelepiped evolves under the dynamics
in the phase space,
and are obtained by examining the rate of stretching of vectors
tangential to the flow. The Lyapunov exponent corresponding to the
$\theta$ freedom where the flow is uniform has the value zero.

To compute the Lyapunov exponent it is necessary to propagate
$k$-dimensional orthonormal vectors, $\hat{\e}^m, m=1,\ldots, k$ in the
tangent space, namely according to the dynamics
\begin{equation}
{\e}_j^m  =   \JF(\X_j,\theta_j)~\cdot \hat{\e}_{j-1}^m
\end{equation}\noindent
where $\JF(\X,\theta)$ is the Jacobian matrix of $\F$ along the orbit
and the subscript $j$ refers to the time step. At each step along the
trajectory, the vectors ${\e}_j^m$ are reorthogonalized, and the norms
give the expansion (or contraction) along the different directions in
phase space [Benettin, Galgani and Strelcyn, 1976]. From this it is
possible to compute the spectrum of Lyapunov exponents as
\beq
\Lambda^m = \lim_{N\to \infty} \frac{1}{N} \sum_{j=1}^N \ln \|{\e}_{j}^m
\|,~~~~~~~~ m = 1,2,\ldots k.
\label{lyap}
\eqn
The quantity  $\ln \|{\e}_{j}^m \| \equiv y_j^m$ is the ``stretch--exponent'':
this measures the expansion (or contraction) factor at step $j$ in the
$m$th direction.
One can further define local, $N$--step, or
finite--time Lyapunov exponents as
\beq
\label{loclyap}
\lambda_N^m  = \frac{1}{N} \sum_{j=1}^N y_j^m,~~~~~ m = 1,2,\ldots k;
\eqn
examining the distribution of these quantities has proven useful in the
study of fractal attractors (see below).

Most studies have focused on the largest of the Lyapunov exponents,
$\Lambda^1$, which is of most importance in determining the dynamics;
the superscript will henceforth be omitted. This is not to suggest
that the other Lyapunov exponents are unimportant: indeed, in higher
dimensional systems with so--called unstable dimensional variability
[Dawson \etl, 1994; Kostelich \etl, 1997] the higher Lyapunov
exponents, $\Lambda^2$, etc. play a major role. (For dynamical systems
described by a set of coupled differential equations, Lyapunov spectra
and finite time Lyapunov exponents can be similarly defined.)

A number of reliable methods are available for the computation
of the asymptotic Lyapunov exponents [Bennettin, Galgani, and Strelcyn,
1976; Eckmann and Ruelle, 1985],
although for experimental data, when working with a single time series,
even this can be a problem since it is often difficult to reliably
compute a small negative Lyapunov exponent using standard algorithms
[Wolf \etl, 1985]. The effect of noise within experimental data can be
removed to some extent by filtering techniques, and recently developed
methods have made it possible to extract negative Lyapunov exponents
from model and experimental data with some degree of success [Huang
\etl, 1994].

Although the largest Lyapunov exponent is negative on SNAs, owing to
the strangeness of the attractor, the dynamics can be {\it locally}
unstable. The negative Lyapunov exponent arises from the fact that
these instabilities are compensated by regions where the dynamics is
locally stable. Such effects are best explored by the examination of
the distribution of the
finite--time Lyapunov exponents [Grassberger \etl, 1988;
Abarbanel \etl, 1991] defined in Eq.~(\ref{loclyap}). The stationary
density of finite--time Lyapunov exponents is
\begin{eqnarray}
\label{prob}
P(N, \lambda) d\lambda & \equiv & \mbox{ Probability that}
~\lambda_N \mbox{~takes a value}\\
\nonumber
& & \mbox{~ between}  ~~\lambda \mbox{~and~} \lambda + d\lambda,
\end{eqnarray}\noindent
and this can be obtained by taking a long trajectory, dividing it in
segments of length $N$, and calculating $\lambda_N$ in each segment.
Since $\lim_{N \to \infty} P(N, \lambda) \to \delta (\Lambda
-\lambda)$, for finite $N$ the density is usually peaked around the
asymptotic \lye, and has a characteristic shape and width [Prasad and
Ramaswamy, 1999].  Several studies [Pikovsky and Feudel, 1995; Prasad
and Ramaswamy, 1998 and 1999] have established that for SNAs the
local instability results in the density of local Lyapunov exponents
having some component at positive $\lambda$; this component decreases
with $N$, and can be used to characterize SNAs. An example of such a
density is shown in Fig.~3 for a SNA in the forced logistic map;
$\Lambda$ is negative, but even for long stretches of any
trajectory---the time interval $N$ = 1000 in this instance---the
dynamics can be unstable, giving a positive local Lyapunov exponent.
This feature distinguishes SNAs from other (periodic or quasiperiodic)
nonchaotic attractors.

The fractality of the attractor cannot be as easily determined through
computation of the fractal dimensions such as the capacity ($D_0$) and
information ($D_1$) dimensions using standard algorithms [Ding \etl,
1989]. The Kaplan--Yorke conjecture [Kaplan and Yorke, 1979] for the
Lyapunov dimension $D_L$, if expected to be valid even in this
situation, gives the estimate $D_L = D_1 =1$; this distinguishes SNAs
from very similar chaotic attractors wherein $D_1$ would be expected to
be larger than 1.  Because of conflicting local and global stability
features on SNAs, numerical results are not always conclusive: an
inordinately large number of points is needed to get converged
exponents for fractal dimensions [Ding \etl, 1989].

Although the criteria discussed above, based on the behaviour of
Lyapunov exponents and fractal dimensions, are essential to a
mathematical characterization of SNAs, other measures are needed,
particularly for examining numerical or experimental data. We discuss
some of these next.

\subsection{Phase and Parameter Sensitivity:}  If the attractor $x(\theta)$ is
viewed as a fractal curve, then its non differentiability can be
detected by examining the separation of points that are initially close
in $\theta$. Pikovsky and Feudel [1995] introduced a measure to
characterize strangeness by calculating the derivative $dx/d\theta$
along an orbit, and finding its maximal value. This yields the phase
sensitivity function
\beq
\Gamma_N =\mbox{min}_{x,\theta}[\mbox{max}_{1<N}\vert dx_N/d\theta
\vert],
\eqn
as the smallest such realization for arbitrary ($x, \theta$), so that a
bound can be set on the rate of growth of $\Gamma$ over the entire
attractor. For a chaotic attractor, the sensitivity grows exponentially
while for a SNA, $\Gamma_N$ (shown in Fig.~4) grows as a power,
\beq
\Gamma_N \propto N^{\mu},
\eqn
with $\mu$ typically $>1$ [Nishikawa and Kaneko, 1996; Pikovsky and
Feudel, 1995]. Other similar
measures can also be devised to distinguish between strange and
non--strange attractors.  For example, if one computes the integral of
the derivative $dx_N/d\theta$ as a function of $\theta$: for a smooth
curve, this integral converges, whereas if there are an infinite number
of discontinuities in $x(\theta)$ as for a fractal, it diverges.

The phase sensitivity seems to be a robust method for determining the
fractality of the attractor [Pikovsky and Feudel, 1995]; this method
can be directly generalized to higher dimensional systems as well
[Sosnovtseva \etl, 1996]. Sensitivity to a parameter, say to the
amplitude of the driving term can be analogously defined [Nishikawa and
Kaneko, 1996], as
\beq
\Gamma_N^{\epsilon} =\mbox{min}_{x,\theta}[\mbox{max}_{1<N}\vert
dx_N/d\epsilon \vert],
\eqn
and this quantity also shows a power--law dependence on $N$.

\subsection{Correlations and Power spectra:} Pikovsky \etl [1995] considered
the averaged squared autocorrelation function as a means of
distinguishing SNA dynamics from chaotic motion. The autocorrelation
function for the dynamical variable is defined in the usual manner,
namely
\beq
C(\tau) = \frac{<x_i x_{i+\tau}>-<x_i><x_{i+\tau}>}{<x_i^2>-<x_i>^2}
\eqn
where $i$ is a discrete time index, $\tau = 0,1,\ldots$ is the time
shift, and $<>$ denotes a time--average. For periodic motion
\beq
C_{av}(t) = \frac{1}{t} \sum_{\tau=0}^{t-1}\vert C(t) \vert^2 ,
\eqn
the average squared autocorrelation asymptotes to 1 ($C(\tau)$
oscillates between 0 and 1). Since quasiperiodic motion does not recur
exactly, for such dynamics the average asymptotes to a value less than
1.  For chaotic motion, $C_{av}$ decreases exponentially to 0, while
for the case of SNAs the behaviour is intermediate.

More quantitative distinctions come from studies of the discrete
Fourier transform of the time series $\{x\}$ generated by the
dynamical equation. The number of peaks in the transform,
\beq
\label{scaling}
T_k = \sum_{m=1}^N x_m \exp(- i 2 \pi m k/N),~~~k = 0,\ldots, N
\eqn
above a threshold $\sigma$ shows the scaling $ {\cal N}(\sigma) \sim
\sigma^{\alpha}$, where the exponent $\alpha$ is between 1 and 2 if the
attractor is strange. This measure can be used to distinguish SNA
dynamics from other nonchaotic behaviour: for motion on an
($n-1$)--frequency quasiperiodic attractor the variation is $ {\cal
N}(\sigma) \sim (1/\log \sigma)^{n-1}$ [Romeiras \etl, 1987].

Other transforms have also been used, as for example the
partial Fourier sums,
\beq
T(\Omega, N) = \sum_{k=1}^N x_k \exp (i2\pi k \Omega),
\eqn
where $\Omega$ is proportional to the irrational driving frequency
$\omega$ [Pikovsky \etl, 1995; Yal\c{c}inkaya and Lai, 1997].  The
graph of Re $T$ vs.  Im $T$ is a curve on the plane which may be
treated as a ``walk'', and one can compute the mean square displacement
$ \vert T(\Omega, N)\vert^2$. If the ``walk'' is Brownian, then $
\vert T(\Omega, N)\vert^2 \sim N$, and the spectrum is continuous,
while if $ \vert T(\Omega, N)\vert^2 \sim N^2$, there is a discrete
spectral component at frequency $\Omega$. SNAs, in contrast, have a
singular--continuous spectrum, which implies the scaling $ \vert
T(\Omega, N) \vert^2 \sim N^{\beta}$ with $1 < \beta < 2$ (see Fig.~5).

\section{SCENARIOS FOR THE FORMATION OF SNAs}

In the absence of quasiperiodic forcing, a dissipative nonlinear
dynamical system typically has periodic attractors or chaotic
attractors. With the introduction of quasiperiodic driving, as for
example in the mapping in Eq.~(\ref{function}) or the flows in
Eqs.~(\ref{zmb}-\ref{heagyditto}), periodic attractors become
quasiperiodic attractors: the motion lies on tori in the phase space,
and these are characterized by the number of independent frequencies
contributing to the quasiperiodicity.  When the parameter itself is
further varied, quasiperiodic or chaotic attractors can become SNAs.

Exactly what constitutes a distinct route or mechanism for the
formation of a strange nonchaotic attractor is somewhat nebulous since
the bifurcations of quasiperiodically driven systems have not been
studied in detail. However, several routes to SNAs have been described
in the literature and there are parallels with the different scenarios
through which chaotic attractors are created [Eckmann, 1981; Ott,
1994]. In the next Sec. we will examine the variation of the
Lyapunov exponent (and its variance) as the control parameter is
varied. Several studies have established that the different scenarios
have distinctive signatures [Prasad \etl, 1998].  For that purpose, it
is helpful to first describe these bifurcations by discussing the
dynamics as a function of a control parameter for several known
scenarios for the formation of SNAs. These are depicted in Fig.~6,
where the torus (before the bifurcation) and the SNA (after the
bifurcation) are shown.

\subsection{Torus Collisions:} One situation wherein the bifurcation is well
studied is the mechanism for SNA formation via torus collisions.

A period--doubling bifurcation in a quasiperiodically driven system
gives rise to a stable doubled torus, with the parent torus becoming
unstable. (Without driving, this is just the pitchfork bifurcation
[Ott, 1994], a period 2$^n$ orbit becomes unstable and bifurcates to a
period 2$^{n+1}$ orbit.)

Heagy and Hammel [1994] identified the birth of a SNA with the
collision between the doubled quasiperiodic torus and its unstable
parent. This requires first that a period--doubling bifurcation occur,
after which the stable torus attractor gets progressively more
``wrinkled'' as the parameters in the system change, i.e., ${\bf
X}(\theta)$ becomes more and more oscillatory as in Fig.~6(a).
Concurrently, the unstable parent torus also gets more oscillatory.  At
an analogue of the attractor--merging crisis that occurs in chaotic
systems [Grebogi \etl, 1987], the SNA is created at the collision of
the period--doubled torus and its unstable parent as in Fig.~6(b), the
Lyapunov exponent remaining negative throughout this crisis.

This route has been seen in a number of different systems, most notably
in the quasiperiodically forced logistic map, Eq.~(\ref{lo}),
which has been extensively studied [Heagy and Hammel, 1994; Prasad
\etl, 1998]. Feudel, Kurths, and Pikovsky [1995] have also identified
other torus collision mechanisms that are operative in forced circle
maps, where a stable and unstable torus intersect on a dense set of
points to give a SNA. A torus collision is thus a general feature of
forced systems and is a common mechanism for SNA creation.

\subsection{Fractalization:} In the ``fractalization'' route for the creation
of SNAs [Kaneko, 1984; Nishikawa and Kaneko, 1996] a quasiperiodic
torus gets increasingly wrinkled and transforms into a SNA without the
apparent mediation of any nearby unstable periodic orbit.
Fractalization is also a likely cause for the interruption of the
cascade of torus doublings [Kaneko, 1984]. It appears unlikely that
there is an explicit bifurcation that is involved in this scenario, but
in some sense this is the most general route to SNA that is observed in
driven systems;  Fig.~6(c,d) shows a torus and its fractalized SNA across
the transition.

\subsection{Intermittency:} The intermittency scenario for the formation of
SNAs [Prasad \etl, 1997; Witt \etl, 1997] is as follows. Upon varying a
parameter, a chaotic strange attractor is first transformed into a SNA
(see Fig.~6(e,f)) which is then eventually replaced by a one-frequency
torus through an analogue of the saddle--node bifurcation. By varying
the parameter one can traverse the transition in the opposite
direction, so the defining characteristic of the bifurcation is the
abrupt change of a torus to a SNA, as will be discussed in detail in
the next Sec.

The intermittent dynamics at this bifurcation are of Type~I [Pomeau and
Manneville, 1980]. Since the SNA occupies a much larger portion of the
phase space as compared to the torus from which it originated, this
transition shares some of the features of a widening crisis in unforced
systems [Grebogi \etl,1987; Mehra and Ramaswamy, 1996]. The
discontinuous change of $\Lambda$ at a saddle--node bifurcation is
softened by the quasiperiodic forcing, and for $\alpha<\alpha_c$, the
Lyapunov exponent shows the scaling [Prasad \etl, 1997]
\beq
\label{powerlaw}
\Lambda-\Lambda_c \sim (\alpha_c - \alpha)^{\mu}.
\eqn
This route to SNA is quite general (this was termed Route C by Witt,
Feudel and Pikovsky, [1997]), and has been seen in a number of driven
maps and in flows, such as the forced Duffing equation, as well
[Venkatesan \etl, 1999].

\subsection{The Blowout bifurcation:} In systems with a symmetric
low--dimensional invariant subspace containing a quasiperiodic torus,
a blowout bifurcation [Ott and Sommerer, 1994] leads to the
formation of a SNA [Yal\c{c}inkaya and Lai, 1996].

Trajectories starting in the invariant subspace, ${\cal S}$, remain in
${\cal S}$.  The Lyapunov exponent $\Lambda$ has two components, one of
which, $\Lambda_T$, is defined for trajectories in ${\cal S}$ with
respect to perturbations in a transverse subspace ${\cal T}$. A
positive $\Lambda_T$ indicates that trajectories in the vicinity of
${\cal S}$ are repelled away from it, and this gives rise to
strangeness. At the blowout bifurcation, $\Lambda_T$ changes its sign,
becoming positive as a system parameter varies. If, concurrently,
$\Lambda < 0$, the attractor is a SNA.

Yal\c{c}inkaya and Lai [1996] studied a number of systems wherein
the blowout bifurcation occurs, including the mapping
\beq
F_{a,b}(x_n,\theta_n) = (\alpha \cos 2 \pi \theta_n + b) \sin 2 \pi x_n
\label{lai}
\eqn
and showed the presence of SNAs in a range of parameter values when
$\Lambda_T > 0$ and $\Lambda<0$. The quasiperiodic torus and the
corresponding SNA after the bifurcation are shown in Fig.~6(g,h).  This
route is found only in those situations where an invariant subspace
exists. Although it does not occur, for example, in the driven logistic
map, Eq.~(\ref{lo}), this scenario is quite general and has been seen
in a continuous dynamical system similar to Eq.~(\ref{zmb}).
The system where SNAs were first discovered, namely Eq.~(\ref{gopy}) is
also of this type, the invariant subspace being the line $x=0$.

\subsection{Quasiperiodic Routes:}
Other than the above four principal scenarios,
a number of different quasiperiodic routes to SNAs have been discussed
in the literature.  These scenarios appear in systems wherein the
(unforced) dynamics itself shows quasiperiodicity, as for example the
forced circle map, where
\beq
F_{K,V,\epsilon}(x_n,\theta_n) = x_n + 2 \pi K + V \sin x_n + \epsilon
\cos \theta_n,
\eqn
$K, V$ being parameters, and $\epsilon$ being the forcing amplitude
[Ding \etl, 1989].
These routes are more descriptive of the different dynamical states
that the system passes through (say from an $n$--frequency quasiperiodic
torus attractor to a SNA) than of distinctive bifurcation routes: one
finds that the actual transition to SNAs is through the mechanisms
discussed above. The torus collision scenario is operative in slightly
different form here as well [Feudel \etl, 1995], the collision here is
between a stable and unstable torus at a dense set of points.  We have
also found that the fractalization route also operates in the same
regions.  Since there is an immense variety in the different types of
periodic, quasiperiodic and chaotic dynamical states that are possible
in nonlinear systems, there are likely to be several such states as
precursors to the eventual SNA [Venkatesan and Lakshmanan, 1998].

\subsection{ Homoclinic Collision:} An important class of bifurcations
which are peculiar to driven maps of Harper type, namely Eq.~(\ref{ks})
and its generalizations [Negi and Ramaswamy, 2000b] are homoclinic
collisions leading to the formation of SNAs [Prasad \etl, 1999].
Below the transition, the dynamics is on invariant curves with
a finite number of branches, see Fig.~6(i). As the parameter is
increased, these branches approach each other (the distance between
branches decreases as a power in the effective parameter), eventually
colliding at a dense set of points and
forming an SNA (Fig.~6(j)). It can be shown that the
collision of the invariant curve with itself is accompanied by
an unusual ``symmetry--breaking'' [Prasad \etl, 1999] which also
help to establish the non-positivity of the Lyapunov exponent
on the SNA.
\section{DYNAMICAL TRANSITIONS}

The strange nonchaotic state is dynamically distinct from the strange
chaotic state, and morphologically distinct from the quasiperiodic (and
nonchaotic) torus attractor. The transitions between these different
states as a parameter is varied are quite distinctive, and have been
the focus of considerable interest. In addition to the scenarios for
the creation of SNAs discussed in the previous section, there can be
other bifurcations and crises in these systems.

The Lyapunov exponent is a good order parameter to study these
dynamical transitions. Further, since the attractor changes drastically
at the transition, the variance of the distribution of finite--time
Lyapunov exponents, Eq.~(\ref{prob}), are also known [Prasad \etl,
1998] to provide a good order parameter whereby the
studied.

The different scenarios for the formation of SNAs all have
characteristic signatures in the manner in which the exponent changes
at the transition, and in the manner in which the variance changes at
the transition. Figure~7 shows the behaviour of these quantities for the
bifurcations discussed in Sec. III.

When the transition occurs via torus collisions (the Heagy--Hammel
mechanism, for instance), the Lyapunov exponent typically shows a point
of inflexion while the variance increases (see Fig.~7(a,b)). In the
fractalized route there is no apparent crisis involved, and therefore
the \lye~ and variance increase only slowly as shown in Fig.~7(c,d).
At the saddle--node bifurcation leading to the intermittent SNA, the
Lyapunov exponent shows an abrupt change, with a power--law
(Eq.~(\ref{powerlaw})) dependence
on the parameter on the SNA side of the transition (see Fig.~7(e,f)).
The fluctuations in the Lyapunov exponent (determined, \eg,~ from
considering a large number of trajectories) shows a remarkable and
abrupt increase at the transition.  For the blowout bifurcation, the
behaviour of the Lyapunov exponent is distinctive. Below the
transition, both $\Lambda$ and the transverse exponent $\Lambda_T$ (see
the discussion in Sec. III) are identical and negative. As the
parameter $a$ in Eq.~(\ref{lai}) is varied, they increase and become
zero at the critical value, $a_c$. After the bifurcation, $\Lambda$ is
negative, but $\Lambda_T$ is positive, leading to locally unstable
motion with a corresponding increase in the fluctuations of the
Lyapunov exponent across the transition; see Fig.~7(g,h). Recent
analysis of finite--time Lyapunov exponents at
this bifurcation has shown that there is a
symmetry--breaking that accompanies the transition to SNAs [Prasad
\etl, 1999]. In the homoclinic transition to SNAs which occurs
in maps of Harper type, the Lyapunov exponent changes from zero
to a negative value. At the transition, the exponent converges
as a power--law rather than exponentially: these SNAs are
critical [Negi and Ramaswamy, 2000c]. The
fluctuations increase across the transition as may be expected
from general considerations, but this is not easy to distinguish
numerically owing to the slow convergence of quantities in the
neighborhood of the transition. Thus the counterparts of Figs.~6(i)
and (j) are not very instructive.

In most systems where there are SNAs, there are a large number of
different dynamical states that are possible and therefore a large
number of different dynamical transitions that can occur. For SNA $\to
$ SNA transitions, as has been described and characterized in the
literature, the Lyapunov exponent is a good order parameter. One
situation where this is {\it not} so is the SNA $\to$ chaotic attractor
transition.  The transition is not accompanied by any major change in
the form or shape of the attractor, and the Lyapunov exponent itself
changes only from being negative to positive, passing {\it linearly}
through zero [Lai, 1996] (see Fig.~8(a)). However, the fluctuations on
the attractor seem to increase substantially, with the variance of the
distribution showing a small but noticeable increase across the
transition (Fig.~8(b)).

\subsection*{Symmetry Breaking}
Several bifurcations in dynamical systems are distinguished by
the fact that the Lyapunov exponent passes through zero: common
examples are the pitchfork and tangent bifurcations in
the absence of driving,  when the slope of the map takes
(absolute) value 1 (see for example, Ott [1994]).

With driving, the scenarios for the transition to SNA are, as
discussed in the previous section, analogus to bifurcations leading
to chaotic states, although the Lyapunov exponent typically
does not pass through the value zero (since it is nonpositive
through the transition). In two of the scenarios, however, it
does: these are the blowout bifurcation route
[Yal\c{c}inkaya and Lai, 1996],  and in the
case of homoclinic collisions leading to SNA [Prasad \etl, 1999].

These transitions are accompanied by a novel symmetry breaking.
The Lyapunov exponent, $\Lambda$ measures the average expansion rate
along a trajectory,
\beq
\Lambda = \lim_{N\to \infty} \frac{1}{N} \sum_{j=1}^N y_j,
\label{lyapx}
\eqn
the $y$'s being the stretch--exponents in the principal direction
(namely $m = 1$ in Eq.~(\ref{lyap})). The stretch exponents are
different at each point of the orbit (since there are no periodic
orbits in such systems). In the sum above, there are several ways
in which the Lyapunov exponent can become zero, but it turns out that
at the blowout bifurcation, there is an exact quasiperiodic symmetry
in the stretch exponents. This can be seen by plotting $y_j$
versus $y_{j+1}$, the return map for stretch exponents
as in Fig.~9(a), where the symmetry with
respect to the diagonal is clearly evident. This holds exactly
at the bifurcation point here as well as in the case of homoclinic
collisions in the Harper map [Prasad \etl, 1999]. When the symmetry
is broken, the total Lyapunov exponent becomes negative and the dynamics
is on a SNA: see Fig.~9(b).

There are other possibilities that will result in a zero Lyapunov
exponent. The symmetries may be more complex [Prasad \etl, 2000],
for instance, or there may be no symmetries at all, as in the SNA
$\to$ chaos transition [Negi \etl, 2000].

\section{EXPERIMENTS~AND APPLICATIONS}

Given the ubiquity of SNA dynamics in quasiperiodically driven systems,
one of the main issues with respect to the experimental observation of
SNAs is whether such attractors are robust to noise.

It is known [Crutchfield, Farmer and Huberman, 1982] that noise
generally lowers the threshold for chaos: systems with additive
noise have a {\it larger} Lyapunov exponent for smaller
nonlinearity. However, the effect of noise depends on the nature
of the system (see \eg, Schroer \etl [1998] and references
therein, and Prasad and Ramaswamy [2000]). Quasiperiodically
driven systems also show an enhancement in the Lyapunov exponent,
but depending on the system, \MLE in the presence of noise can
still be negative, and the SNAs can continue to exist [Prasad,
Mehra and Ramaswamy, 1998].  The addition of noise usually
``smears out'' the attractors, and the threshold values for
bifurcations typically shift to lower parameter values, but the
actual transitions---now from noisy tori to noisy SNAs---survive.

A number of different experiments have verified the existence of SNAs,
ranging from driven mechanical systems such as the magnetoelastic
ribbon experiment [Ditto \etl, 1990] to electronic circuits.  Zhou
\etl[1992] studied the model SQUID system, Eq.~(\ref{zmb}) above,
on an analog simulator, and verified SNA dynamics by computing power
spectra and Lyapunov experiments. Similar experiments have been
implemented in the forced Ueda's circuit [Liu and Zhu, 1996], the
forced Chua's circuit [Zhu and Liu, 1997] and the forced
Murali--Lakshmanan--Chua circuit [Yang and Bilimgut, 1997].

It should be noted that for the experimental characterization of SNAs,
there are practical problems and limitations. Lyapunov exponents can be
estimated from experimental data, but the inherent noise can make it
very difficult to determine a small negative exponent. The error bounds
obtained by application of standard algorithms can be large enough to
make estimates inconclusive. The same holds for fractal dimension
estimates.

Experimentally, the spectral distribution function measure discussed in
Sec. 2 has been used in addition to the Lyapunov exponent or the
fractal dimension obtained from analysis of time series data. In most
cases, the scaling behaviour of this function is more unambiguous than
other measures.

Plasma systems have also been extensively used in order to study a
number of nonlinear dynamical effects, and in recent measurements
of the glow discharge in a neon gas plasma [Ding \etl, 1997], SNAs have been
observed in the absence of any driving. The quasiperiodic driving comes
from autoexcitations of the plasma, and the nature of the attractor was
verified through dimension and Lyapunov exponent estimations. Another
SNA system where there is no explicit external forcing is the model
study of a neuronal membrane system as well as the EEG data examined by
Mandell and Selz [1993], but the origin of the quasiperiodic driving
there is less clear.

One property of SNAs that make them interesting experimental systems
for study---from the viewpoint of potential applications---is their
relative ease of synchronization. In recent years the synchronization
and control of chaotic systems has been the focus of much research
activity [Shinbrot, 1995; Chen and Dong, 1998]. Two identical nonlinear
chaotic dynamical systems can be made to synchronize by using one (the
{\it drive}) to drive the other (the {\it response}). Pecora and
Carroll [1990] showed that if the Lyapunov exponents corresponding to
the response subsytem were negative, then synchronization would occur.

Such synchronization is trivially achieved with SNAs since the largest
Lyapunov exponent is already negative. Further, even coupling the
systems is unnecessary so long as the initial phases are matched, again
because of the negative Lyapunov exponents [Ramaswamy, 1997]; see
Fig.~1(c) for an illustration.

In juxtaposition with the fact that the dynamics on SNAs is aperiodic,
this gives rise to interesting possibilities for their use. Two recent
proposals outline the use of SNAs in the area of secure communications.
Zhou and Chen [1997] transmit digital information by modulation of a
system parameter so that the dynamics switches between a SNA and
another chaotic or nonchaotic attractor. Information is recovered at
the receiver by employing the synchronization between the nonchaotic
attractor of the receiver and the transmitter. Another implementation
[Ramaswamy, 1997] uses two identical independent SNAs which are
synchronized by in-phase driving but are otherwise uncoupled.  The
principle employed is similar to that of chaotic masking. A
low--amplitude information signal is added to the output of the
transmitter SNA system. Since the SNA dynamics is aperiodic, the
resulting signal also aperiodic. The receiver SNA system is exactly
synchronized with the transmitter, and thus the information can be
simply recovered by subtracting the two signals (see Fig.~3 in
Ramaswamy, [1997]).

\section{SUMMARY}

Strange nonchaotic attractors are an important class of dynamical
attractors that are generic in quasiperiodically driven nonlinear
dynamical systems, both mappings as well as flows. Systems where
SNAs arise naturally span a wide range since the possibility of
such dynamics devolves on a combination of dissipation,
nonlinearity and quasiperiodic modulation.

The skew--product structure (whereby the ``system'' dynamics does
not feed into the dynamics of the forcing term) is common to all
examples of systems with SNAs that have been studied so far.
Furthermore, explicit quasiperiodic driving is also a feature of
hitherto studied systems. However, both these features are probably not
necessary in order that SNAs be formed [Negi and Ramaswamy, 2000a].
In particular, driving a system with signals based on fractal
sequences also appears to yield SNAs [Kuptsov, 1998].

Very recently Cassol, Veiga and Tragtenberg [Cassol \etl, 2000] have
studied a map with {\it periodic} forcing, where they find
apparent strange nonchaotic motion with Lyapunov exponent equal
to zero, and orbits on a fractal attractor. No other examples
of periodically driven systems with SNAs are known so far, and
from general considerations, it would appear unlikely that such
attractors can occur under conditions of periodic forcing.  Anishchenko
\etl [1996] reported the presence of a \sna~ in a periodically forced
system, but this result has been questioned [Pikovsky and Feudel,
1997].  We have also examined this system and find only chaotic and
periodic attractors.

Modulation of chaotic systems can lower the Lyapunov exponent by
enhancing the measure on contracting regions of phase space. This
means of ``control'' can sometimes create SNAs if the exponent is
sufficiently lowered, as discussed recently [Shuai and Wong, 1998,
1999].  For the case of chaotic  or noisy driving, Rajasekhar [1995]
has shown the possibility of nonchaotic dynamics via a type of
chaos control, but this methodology does not necessarily lead to
attractors [Prasad and Ramaswamy, 2000].  The question of whether
there can be SNAs with other types of forcing remains open.

In this review, we have described the general phenomenology of such
systems, with particular emphasis on the scenarios for the creation of
SNAs.  We have also described the many different methods that have been
employed to confirm strange nonchaoticity. SNAs are relevant in a
number of theoretical and experimental situations, and these have been
discussed in some detail. One of the other principal themes in current
research is the use of the Lyapunov exponent and its fluctuations as
order--parameters for characterizing and determining different
dynamical transitions in these systems.  Both the exponent as well as
the fluctuations show characteristic variation with system parameters,
and this is as true of transitions from tori to SNAs as it is for
transitions from SNAs to SNAs, or from SNAs to chaotic attractors.

Applications and experiments that exploit the unusual properties of
SNAs are beginning to be devised.  Much of our discussion has centered
around the dynamics of low--dimensional systems since these are simple
enough to analyse and most of the concepts and phenomena in the study
of SNAs can be illustrated here. Extensions to higher dimensions of
many of the phenomena and arguments are straightforward, and have begun
to be explored in the literature.

\newpage
\section*{\sc Acknowledgment} We are grateful to Mike Cross and
Jim Heagy for discussions and for their comments on an earlier
version of this manuscript. We would also like to thank Jim
Heagy for sharing his code for locating unstable tori, and M
Lakshmanan and A Venkatesan and Vishal Mehra for discussions
on SNAs.  This work was supported by a grant from the Department
of Science and Technology, India.

\newpage
\centerline{Figure Captions}

\begin{itemize}

\item[Fig.~1.] The trajectory of a strange (a) nonchaotic  and (b)
chaotic attractor of Eq.~(\ref{zmb}) at $q_2=0.88$~ and $q_2=0.38$
respectively. The other parameters are $k=\beta=2$ and $q_1=2.768$.
In c) and d), trajectories with two different initial conditions, $x$
(dotted line) and $x^{\prime}$ (solid line) are shown for the
attractors of a) and b) respectively.

\item[Fig. 2] Schematic phase diagram for the forced logistic map,
 Eq.~(\ref{lo}). The
rescaled parameter $\epsilon'$ is defined as $\epsilon^{\prime} =
\epsilon/(4/\alpha-1)$.  P and C correspond to regions of
torus and chaotic attractors.  SNAs are mainly found in the shaded
region along the boundary of P and C (marked S).  The actual boundaries
separating the different regions are more convoluted than shown, and
regions of SNA and chaotic attractors are interwoven in a complicated
manner.  Intermittent SNAs are found on the edge of the C$_2$ region
marked I, while the left boundary of C$_2$ has only fractalized SNAs.
Along the boundary of C$_1$, SNAs formed by either the torus collision
mechanism or fractalization can be found. This phase diagram is
discussed in detail in Prasad \etl [1997, 1998].

\item[Fig.~3.] Density of finite--time Lyapunov exponents $P(1000,
\lambda)$ for an intermittent SNA in the forced logistic map,
Eq.~(\ref{lo}) at $\alpha = 3.84549, \epsilon^{\prime} = 0.073$. See
Prasad and Ramaswamy [1998] for details.

\item[Fig.~4.] Plot of  the phase sensitivity function $\Gamma_N$
vs $N$ in the logistic map, Eq.~(\ref{lo}) along $\epsilon^{\prime}
=1$, for the fractalized \sna~ at $\alpha=2.7$ (solid line) and a
nearby torus at $\alpha=2.5$ (dotted line). The exponent was found to
be $\approx 2.35$ (the dashed line is a least--squares fit).
Note that this numerical value of the exponent depends on the system
and its parameters; but in all cases $\Gamma_N$ diverges with $N$.

\item[Fig.~5.] $\vert T(\Omega, N) \vert^2$, the singular--continuous
spectral component, plotted versus $N$ for an intermittent \sna~ at
$\epsilon^{\prime} =1$, $\alpha=3.405805$ in the quasiperiodic logistic
map,  Eq.~(\ref{lo}) for $\Omega \equiv \omega/4$.
 The measured slope, here $\approx
1.75$ (dashed line), changes with parameters and from system to system.

\item[Fig.~6.] A sequence of torus attractors and the corresponding
\sna~ at the transition (indicated by the vertical dotted line in Fig.~7)
in different scenarios.  In the forced logistic map, Eq.~(\ref{lo})  with
$\epsilon^{\prime} = 0.3$, a) the tori at $ \alpha=3.487$ and b) the
SNA at $\alpha=3.489$ following the Heagy--Hammel route.  The dashed
line in a) and b) is the unstable period-1 repeller.  Fractalization at
$\epsilon^{\prime} =1 $ c) the torus attractor at $ \alpha=2.6$ and d)
the SNA at $\alpha=2.7$.  The intermittency transition at
$\epsilon^{\prime} =1$, with e) the wrinkled torus at $\alpha=3.405809$
and f) the SNA at $\alpha=3.405808$.  The example of the blowout
bifurcation route to SNA is in the mapping Eq.~(\ref{lai}) with
parameter $b = 1$; g) the torus at $a=1.9$ and h) the SNA at $a=2.1$.
The homoclinic collision route in the Harper map, Eq.~(\ref{ks})
with parameters $E = 0$, $\omega$ the golden mean ratio; i) the invariant
curves at $\alpha = 0.8$, below the transition, and j) the SNA at
$\alpha = 1.08$. In (i) and (j), for convenience the variable plotted
on the ordinate is tanh($x$), which has the range [-1,1]  rather than $x$
which has the range ($-\infty, \infty$).

\item[Fig.~7.] Plot of the largest \lye~ (left) and its variance (right)
across the transition from tori to SNAs, corresponding to the plots
  shown in Fig.~6.
The vertical line in each panel is drawn at the parameter value
corresponding to the transition.  (a)
and (b) are for the Heagy-Hammel route in the logistic map, Eq.~(\ref{lo})
  at the
transition at $\alpha_c = 3.48 779 \ldots$ for $\epsilon^{\prime}
=0.3$; (c) and (d) are for the fractalization route, also in the same
system, with $\alpha_c =2.6526 \ldots$ at  $\epsilon^{\prime} =1$;
(e) and (f) are for the intermittent transition at $\alpha_c
=3.405808806\ldots$, $\epsilon^{\prime} =1$ and (g) and (h) are for the
blowout bifurcation route in the mapping, Eq.~(\ref{lai}), with $\a_c
=2.0$ at $b=1$.

\item[Fig.~8.] The transition from SNA to a chaotic attractor in the
logistic map, Eq.~(\ref{lo}) along
$\epsilon^{\prime}=0.3$. (a) The Lyapunov exponent across the
transition, and (b) its fluctuations, $\sigma$, calculated from $50$
samples, each of total length $10^7$ iterations.

\item[Fig.~9.] The symmetry breaking in the blowout route to
SNA. The return map for the stretch--exponents (a) at
the bifurcation point in the mapping, Eq.~(\ref{lai}),
with $\a_c =2.0$ at $b=1$, and (b) at $\alpha = 2.01$, when the
dynamics is on a SNA and the symmetry is clearly broken.

\end{itemize}
\end{document}